\documentclass[]{aa}
\usepackage{natbib,psfig,graphicx,ulem}
\begin{document}

% \thesaurus{02.07.1; 03.13.8; 11.03.4;
% 11.05.1; 11.05.2; 11.06.1; 11.06.2; 11.19.6}

\title{Deep and wide
field imaging of the Coma cluster: the data\thanks{Based on
observations obtained at the Canada-France-Hawaii Telescope (CFHT)
which is operated by the National Research Council of Canada, the
Institut des Sciences de l'Univers of the Centre National de la
Recherche Scientifique and the University of Hawaii.}}

%\offprints{C. Adami \email{christophe.adami@oamp.fr}}

\author{C. Adami\inst{1} \and J.P.
Picat\inst{2} \and C. Savine\inst{1} \and A. Mazure\inst{1} \and
M.J. West\inst{3} \and J.C. Cuillandre\inst{4} \and R.
Pell\'o\inst{2} \and A. Biviano\inst{5} \and C.J.
Conselice\inst{6} \and F. Durret\inst{7,8} \and J.S. Gallagher
III\inst{9}  \and M. Gregg\inst{10} \and C. Moreau\inst{1} \and 
M. Ulmer\inst{11} }

\institute{ LAM, Traverse du Siphon, 13012 Marseille, France \and
Observatoire Midi-Pyr\'en\'ees, 14 Av. Edouard Belin, 31400
Toulouse, France \and Department of Physics and Astronomy,
University of Hawaii, 200 West Kawili Street, Hilo HI
96720-4091, USA \and Canada-France-Hawaii Telescope Corporation,
65-1238 Mamalahoa Highway, Kamuela, HI 96743 \and INAF/Osservatorio
Astronomico di Trieste, via G.B. Tiepolo, 11, 34131 Trieste, Italy
\and Department of Astronomy, Caltech, MS 105-24, Pasadena CA
91125, USA \and Institut d'Astrophysique de Paris, CNRS,
Universit\'e Pierre et Marie Curie, 98bis Bd Arago, 75014 Paris,
France \and Observatoire de Paris, LERMA, 61 Av. de
l'Observatoire, 75014 Paris, France \and University of Wisconsin,
Department of Astronomy, 475 N. Charter St., Madison, WI 53706,
USA \and Department of Physics, University of California at Davis,
1 Shields Avenue, Davis, CA 95616 \and Northwestern University, 2131
Sheridan, 60208-2900 Evanston, USA }

\date{Accepted . Received ; Draft printed: \today}

\authorrunning{Adami et al.}
\titlerunning{Deep and wide field imaging of the Coma cluster: the
data}

\maketitle

\abstract {We have obtained deep and wide field imaging of the Coma
cluster of galaxies with the CFH12K camera at CFHT in the B, V, R and
I filters. In this paper, we present the observations, data reduction,
catalogs and first scientific results.

We investigated the quality of our data by internal and external literature
comparisons. We also checked the realisation of the observational requirements
we set.

Our observations cover two partially overlapping areas of
$42 \times 28$ arcmin$^2$, leading to a total area of 0.72 $\times$
0.82 deg$^2$.
We have produced catalogs of objects that
cover a range of more than 10 magnitudes and are complete
at the 90$\%$ level at B$\sim$25, V$\sim$24, R$\sim$24
and I$\sim$23.5 for stellar-like objects, and at B$\sim$22, V$\sim$21,
R$\sim$20.75 and I$\sim$20.5 for faint low-surface-brightness galaxy-like
objects.
Magnitudes are in good agreement with published values from R$\sim$16
to R$\sim$25. The photometric uncertainties are of the order of 0.1
magnitude at
R$\sim$20 and of 0.3 magnitude at R$\sim$25.
Astrometry is accurate
to 0.5~arcsec and also in good agreement with published data.

Our catalog provides a rich dataset that can be mined
for years to come to gain new insights into the formation and
evolution of the Coma cluster and its galaxy population.
As an illustration of the data quality, we examine the
bright part of the Colour Magnitude Relation (B-R versus R) 
derived from the catalog and find that it is in
excellent agreement
with that derived for
galaxies with redshifts in the Coma cluster, and with previous CMRs
estimated in the literature.

}


\section{Introduction}\label{sec:intro}

Long considered the archetype of rich and relaxed
clusters, Coma is now known to have a rather complex structure, with a
number of substructures detected at various wavelengths (see Biviano
1998 for a complete review on the Coma cluster up to 1995).
Indeed,
although a large quantity of multi-wavelength observational
 data are presently available
for Coma, our theoretical understanding of this unique cluster is
still far from complete.

Previous photometric studies of the Coma cluster
were done in few colors, and were
limited either to wide but shallow surveys or to small deep ones, often
obtained under less-than-ideal seeing conditions.
For example, very large shallow surveys (e.g. Godwin \& Peach
1977, hereafter GP77) were the basis for important studies of the Coma
cluster luminosity function and of the properties of galaxies
in general.
However, due to the shallowness of this catalog, very
little was known about Coma's galaxy population at magnitudes
fainter than R$\simeq$20.  Recent deeper catalogs are
still limited to R$\sim$21 when covering the whole cluster
(e.g. Castander et al. 2001, Terlevich et al. 2001 or Komiyama et
al. 2002). Finally, very deep catalogs such as that of Bernstein et
al.(1995) covered only a very small area of sky.   While many of these
previous studies led to important discoveries, all suffered to
some extent from insufficient depth or coverage, which
made it difficult to generalize the results obtained from
these studies.

Here we report a new catalog of Coma cluster galaxies,
obtained with
the CFH12K camera
 (Cuillandre et al.  2000) installed at the
prime focus of the Canada-France-Hawaii Telescope (CFHT),
which provides
the first deep, wide-field images of Coma in the B, V, R and I filters, in
seeing conditions well below 1~arcsec for a large part of the
observations. As the field covered by the CFH12K is 0.32 deg$^2$ we
made two overlapping pointings to cover a total field of 0.6
deg$^2$. The data obtained are significantly deeper than any of the
recent CCD large field imaging of the Coma cluster; e.g. Terlevich et
al. 2001 (U and V filters, depth V=20, 1~deg$^2$, 2.2~arcsec seeing)
or Komiyama et al. 2002 (B and R filters, 2.25~deg$^2$, complete to
R=21, seeing from 0.8 to 1.5~arcsec). A BVR trichromic image of the final
result is shown in Fig. ~\ref{pub}.

\begin{figure*}
%\centering \mbox{\psfig{figure=3810f1.ps,height=21.5cm,angle=0}}
\caption[]{BVR color image of the Coma cluster from our CFH12K data. The size
  of the field is 0.72deg$\times$0.82deg. North is up and East is to the left.}
\label{pub}
\end{figure*}

Our main scientific goals in obtaining deep wide-field multi-color images of
Coma
are: a)~to study environmental effects on the faint end slope of the
galaxy luminosity function (as a very significant extension of, e.g.,
the work by Lobo et al. 1997 or by Andreon $\&$ Cuillandre 2002);
b)~to analyze Coma's morphological and dynamical structure to
see how
it relates to the cluster's formation history
(e.g. Mellier et al. 1988
or Adami et al. 2005b), the diffuse light emission (Adami et
al. 2005a) and the faint and low surface brightness galaxy
distribution (Adami et al., 2006, submitted).
However, because this new Coma catalog provides a rich source of
data that will undoubtedly be useful for other researchers,
we intend to make it publicly available.

In this paper, we describe our data acquisition, reduction, and
cataloguing. Catalogs are available at http://cencosw.oamp.fr/.

\section{Observations and data reduction}

\subsection{Requirements}

Several considerations guided our
survey design; we summarize these below.

Requirement 1: One of our primary science drivers was to
investigate environmental effects on the galaxy population in the
Coma cluster. Limiting ourselves to the densest part of the cluster, we
wanted to cover at least 5 core radii, in order to sample
regions where
the density falls below $\sim$1$\%$ of the central cluster galaxy
density. To do this requires
surveying a region
50$\times$50 arcmin$^2$, which is possible with two adjacent CFH12K fields
(see next section).

Requirement 2: We study the spectral energy distributions of galaxies
in order to constrain galaxy formation models. Four photometric bands
were required for this, in order to obtain color-color plots. We also
wanted to compute photometric redshifts and this requires, at the
cluster redshift, to have at least one band below the 4000\AA\ break
and one above it. For Coma, this means that we needed U band data (not
yet acquired) and B, V, R and I band data (all observed).

Requirement 3: We wanted to sample the dwarf galaxy population as deep
as possible in at least one band, without suffering too much contamination
from
globular clusters. At the redshift of Coma, this implies needing a
complete catalog down to R=24 (see Bernstein et al. 1995).

\subsection{Instrumental setup}

Our data were acquired at the CFHT 3.6m telescope with the CFH12K
CCD mosaic (12 individual 2048$\times$4096 CCDs) (see Cuillandre
et al. 2000 and Table ~\ref{tab:det}) installed at the prime
focus.  The pixel size is 0.206~arcsec, well suited to the mean
seeing at prime focus ($\sim$0.8~arcsec), giving a full field of
view of 42$\times$28~arcmin$^2$ per field.  Two adjacent fields
offset in the North/South direction with a $\sim$7~arcmin overlap
were observed in the B, V, R and I CFH12K filters giving a final
field of view of 0.72$\times$0.82 deg$^2$ (in good agreement with
Requirement 1). The V, R and I are Mould filters while B is a
CFH12K customized filter (see Table ~\ref{tab:filt}). We will
refer hereafter to the filters using the capital letters B, V, R
and I and to the fields as North for the North pointing and South
for the South pointing. These four bands partially satisfy
Requirement 2, and we still plan to acquire the missing U band
data.

\begin{table}
\caption{Main characterictics of the detectors}
\begin{tabular}{ll}
\hline
Detector & CFH12K \\
\hline
Pixel size & 0.206~arcsec \\
Gain & 1.4-1.8 e$^{-}$/ADU (typically 1.5)\\
Quantum efficiency & 80\% at [600,700]nm \\
Readout noise & 3 to 10 e$^{-}$ (typically 5)\\
Dark current & 1 e$^{-}$ per min\\
\hline
\end{tabular}
\label{tab:det}
\end{table}

\begin{table}
\caption{Filter characterictics}
\begin{tabular}{llll}
\hline
Filter & $\lambda$(\AA) & $\Delta$$\lambda$(\AA) & Peak transmission \\
\hline
B & 4310 & ~950 & 85\% \\
V & 5370 & ~940 & 90\% \\
R & 6580 & 1300 & 85\% \\
I & 8223 & 2164 & 91\% \\
\hline
\end{tabular}
\label{tab:filt}
\end{table}

\subsection{Exposure times and first/second epoch data}

A first set of B, V, R observations was obtained in 1999 and 2000
with only 10 of the 12 CCDs available at that time (first epoch
data). A second set of CFH12K images (second epoch data) with all
CCDs working, was obtained in 2000: the South field only in R,
both North and South fields in I. In both runs, long exposures (1
to 2h) were split into several shorter ones, offset by a few
arcsec, in order to improve the cosmetics, flat field and fringing
corrections. Offsets between exposures were kept small enough so
that images could be stacked without correcting for the
differential distortion between exposures.

Exposure times were 24 min in B, 12 to 14 min in V, 10 to 12 min
in R and 12 min in I, rather long (which resulted in bright
galaxies being saturated) in order to be sky background limited
and to minimize the overheads. Tables ~\ref{tab:det} ,
~\ref{tab:filt}, ~\ref{tab:obsfirst} and ~\ref{tab:obssecond},
summarize the main observational informations.

\begin{table*}
\caption{First epoch observations}
\begin{tabular}{lllllll}
\hline Filter & Field & Center (J2000) & Exposures & Total
exposure time &
Mean seeing & Observing year \\
       & & & & (s) & (arcsec) & \\
\hline
B & South & 12:59:40.00 ; 27:48:47.1 & 5 & 7200 & 1.07 & 2000 \\
B & North & 12:59:40.00 ; 28:10:07.0 & 5 & 7200 & 1.01 & 2000 \\
V & South & 12:59:40.00 ; 27:48:47.1 & 5 & 4200 & 1.00 & 2000 \\
V & North & 12:59:40.00 ; 28:10:07.0 & 7 & 5040 & 0.97 & 2000 \\
R & South & 12:59:40.00 ; 27:49:07.2 & 6 & 3300 & 0.90 & 1999 \\
R & North & 12:59:40.00 ; 28:10:06.9 & 5 & 3600 & 0.87 & 2000 \\
\hline
\end{tabular}
\label{tab:obsfirst}
\end{table*}

\begin{table*}
\caption{Second epoch observations}
\begin{tabular}{lllllll}
\hline Filter & Field & Center (J2000) & Exposures & Total
exposure time &
Mean seeing & Observing year \\
       & & & & (s) & (arcsec) & \\
\hline
R & South & 12:59:40.00 ; 27:49:17.0 & 10 & 7200 & 0.85 & 2000 \\
I & South & 12:59:40.00 ; 27:48:46.9 & 10 & 7200 & 0.80 & 2000 \\
I & North & 12:59:40.00 ; 28:10:06.9 & 10 & 7200 & 1.00 & 2000 \\
\hline
\end{tabular}
\label{tab:obssecond}
\end{table*}

\subsection{Data reduction  }

Acquiring and reducing the data was part of C. Savine's PhD Thesis
(Savine 2002) and the process is fully described in her thesis
report (available at http://cencosw.oamp.fr/).
Here we summarize the key steps.

First epoch data were preprocessed by J.C. Cuillandre using his
own  FITS Large Images Processing Software (FLIPS) package, which
performs  functions similar to IRAF task MSCRED and is optimised
to handle large mosaics.

Second epoch data were fully preprocessed at the TERAPIX
(http://terapix.iap.fr/soft/) center using FLIPS and then
processed with the TERAPIX modules for photometric and astrometric
calibrations and mosaic image construction.

In both cases, a `superflat'' built from science images on
``empty'' fields obtained during the same period was used to
subtract the fringe patterns. The mosaic is normalised on CCD 04,
with the higher sky value. Residual variations are very small and
the flat-fielding appears to be good to better than 0.1$\%$ in B
and V across the full mosaic (Kalirai et al. 2001).

 Astrometric and photometric calibration and
mosaic image construction were performed for each CCD chip using
IRAF routines for first epoch data. The APM catalog
(http://www.ast.cam.ac.uk/apm-cat/) has been taken as reference to
transform the distorted (x,y) positions into corrected
($\alpha$,$\delta$) coordinates using the IRAF package
``IMAGE.IMCOORDS''. For second epoch data Terapix modules were
used. With Terapix, photometric and astrometric calibrations are
determined using the entire mosaic to give more homogeneous
solutions. A global astrometric solution for all CCD images
projected onto a common system was computed with comparison to the
USNO catalog for absolute calibration.

Photometric calibration was performed by comparison to a set of
several Landolt fields observed at the same epoch. For each
filter, the zero point determination was found to be quite similar
for the 12 individual CCDs, with a mean error less than 0.025
magnitude, except for the South first epoch data in the R filter
which will be not used in the catalog.

Because the shifts between exposures are small and there are few
exposures, gaps between individual CCDs are only partially
corrected  and present a lower signal-to-noise ratio. However,
this affects only a very small fraction of the surveyed area, and
should have little or no effect on  most studies based on the
catalog.

\section{The catalog}

\subsection{Choice of first/second epoch and North/South data}

In the final catalog, data for objects with declinations $\delta \leq
+28.05^{\circ}$
are taken from the South field, while objects with $\delta \geq
+28.05^{\circ}$ are taken from North data.

On the whole field, the B and V data are only available from first
epoch data and the I band data from second epoch data. The R band
data are taken from the first epoch in the North and from the
second epoch in the South.

\subsection{Combination of B, V, R and I data}

The BVR catalog for first epoch data was built with reference to
the R positions, selecting the nearest object in the other
filters, and checking the magnitude consistency. As the
astrometric solution was determined CCD by CCD using a low order
polynomial function, residual distortions can be large (1 to
3~arcsec) in the corners and at the edges of the individual CCDs,
and the proximity criterion is no longer valid for combination
with data which do not suffer from the same errors. This is the
case for the second epoch data for which a better global
astrometric solution on each of the South and North fields was
derived.

In order to recover the distorted regions, we developed a more
sophisticated geometrical method, searching for the association
which minimizes the distances of all objects in a given area
around the reference objects. This is a pattern recognition
process which
is quasi-independent of residual distortions if the area is small
compared to the distortion scale. In cases when there were
fewer than 2 objects
in common in the exploration area, the nearest object was selected.
The best combinations (in terms of errors and number of matched
objects), confirmed with a map of the associated positions, was
obtained with an exploration area of radius 15~arcsec, allowing for
position errors as large as 3~arcsec. At the end of the process, results
were checked for possible duplicate associations and corrected.

The final result is a single catalog of more than 60,000 objects
(galaxies and point sources) with B, V, R and I data in the
area covered by
the South and North pointings. The object positions
were selected from the R data, from the first epoch in the North and
the second epoch from the South.

\subsection{Catalog content}

The following quantities are given in the catalog:

  - identification number for each object

  - ($\alpha$,$\delta$) coordinates in J2000

  - aperture magnitudes in the 4 bands for a 3~arcsec aperture

  - SExtractor error estimates on these magnitudes

  - total magnitudes in the 4 bands

  - SExtractor error estimates on these magnitudes

  - SExtractor central surface brightness (in the I and R bands)

  - external magnitude errors in the four bands and for each
    measured magnitude

  - shape parameters: minor axis, major axis, orientation (in the I
    and R bands)

  - SExtractor star-galaxy classification (in the I and R bands)

  - our star-galaxy classification in the I band (distinct from the
SExtractor one)

  - SExtractor detection flags (see Bertin $\&$ Arnouts 1996)

  - B, V, and R magnitudes taken from the literature, scaled to our system for
  objects brighter than R=17.5

How these parameters were computed is described below.

\section{Catalog analysis}

\subsection{Object extraction}
Sources were extracted using the SExtractor package (Bertin $\&$
Arnouts 1996). The detection threshold was set to 2$\sigma$, a trade
off between detecting as many objects as possible and limiting the
number of false detections. The minimum number of contiguous pixels above
the detection threshold for extraction was set to 9, a conservative value which
corresponds to about half of the pixels at FWHM for a 0.9~arcsec
seeing. In this way, although we limit our detection of the large
surface brightness population, we increase the confidence in the
detections.

Aperture magnitudes were calculated, in all filters, using the
SExtractor MAGAuto parameter based on the Kron total magnitude.
Central surface brightnesses were computed in the best quality bands
(R and I) only.

\subsection{Internal astrometry}

The astrometry  was computed on the R band images with an
accuracy of about 0.2-0.3 arcsec over the whole field, which is
typical of the precision of the reference catalogs, except in the
small bands at the edges of the CCDs for the first epoch
data.

The relative astrometry with the other filters is much better; for
example, in the B band, the residual dispersion is 0.07 arcsec in
$\alpha$ and 0.08 arcsec in $\delta$.

Comparison of the astrometric solutions between the two epochs
 in the common South field and in the R filter shows an agreement at the
0.5~arcsec level (see Fig. ~\ref{astrom}) for all magnitudes.

\subsection{Astrometry comparison to published data}

For sources brighter than R=20, the comparison of our catalog positions
to those in the USNO and GP77 catalogs
shows a dispersion of about 0.75~arcsec with USNO and 0.9~arcsec with
GP77 in both $\alpha$ and $\delta$. The astrometry dispersion between
our catalog and GP77 is typical of the astrometry error of the GP77
catalog itself.

For sources fainter than R=20, the comparison of our catalog with that of Bernstein et
al. (1995) shows a dispersion of about 0.7~arcsec.

\begin{figure}
\centering \mbox{\psfig{figure=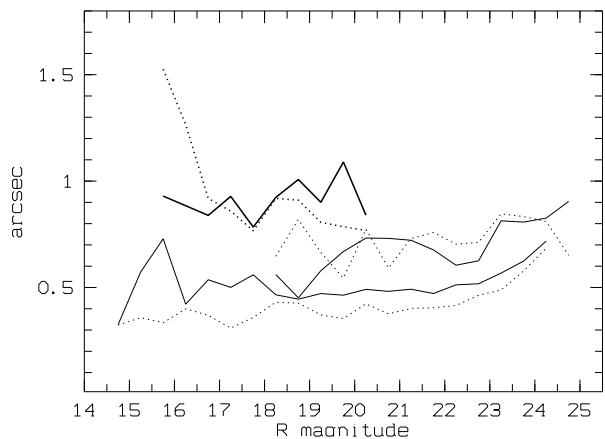,height=6cm,angle=-90}}
\caption[]{Variation of the astrometric dispersion in $\alpha$ (solid
lines) and $\delta$ (dashed lines) as a function of R magnitude. From
bottom to top: Very thin lines (first two lines): comparison between first
and second epoch data. Thin lines (third and fourth lines): comparison with
Bernstein et al. (1995). Thick lines (fifth and sixth lines): comparison
with GP77.}
\label{astrom}
\end{figure}

Results are presented in Fig.~\ref{astrom}.

\subsection{Photometric corrections}

Zero points were computed using observations of standard stars
(Landolt 1992) in the fields SA101, SA104 and SA110. These
standards were observed at nearly the same air mass as the science
observations in the case of first epoch data and scaled to zero
air mass for the second epoch. The correction in magnitude due to
small airmass variations would have been, at maximum, of 0.033
magnitude for R, 0.020 magnitude for V, and 0.015 magnitude for B,
much less than the estimated error in the magnitudes, and hence
were neglected. We also computed central surface brightnesses in
the R and I bands. The calibration was done on the entire mosaic
after scaling each chip, assuming identical color equations for
all CCDs. This proves to be true at better than 4$\%$ for all the
CFH12K observations (McCracken et al. 2003). The B, V and R
filters appear to have negligible color terms with respect to the
Johnson Kron Cousins system and we chose to stay in the CFH12K
system, applying no correction in the catalog for the color
equations, even in I where the correction could be of the order of
I=0.1. Instrumental magnitudes are converted to the Vega system.

We correct for Galactic extinction using the Schlegel et al. (1998)
maps. This contribution remains small: $\sim$0.05 magnitude for the B
band in the worst case.

\subsection{Magnitude error budget}

Magnitude errors are due to zero point uncertainties, seeing
inhomogeneities and internal errors.
These are discussed below.

\subsubsection{Computed error budget}

Zero point uncertainties are smaller than 0.03 magnitude.  In order
to check the effects of seeing conditions on the
photometry, we compared the aperture magnitudes on R
images (0.8~arcsec seeing) to the aperture magnitudes on the same
images convolved with a Gaussian to mimic 1.00~arcsec seeing. We
confirm the results by Saglia et al (1993): when the seeing gets
worse, the objects appear fainter since part of the signal is
scattered into the background. The worst error remains smaller than 0.25
magnitudes at R$\sim$24. We note, however, that such a process
slightly underestimates the error because of the background smoothing.

Internal SExtractor errors are of the order 0.1 magnitude at
R$\sim$24, a large part probably due to close neighbors as
shown in the section which describes the catalog properties.

In total, we therefore expect errors for the faintest catalogued
objects (R$\sim$24) to be of the order of $\sim$0.27 magnitude.

\subsubsection{Externally estimated error budget}

\begin{figure}
\centering \mbox{\psfig{figure=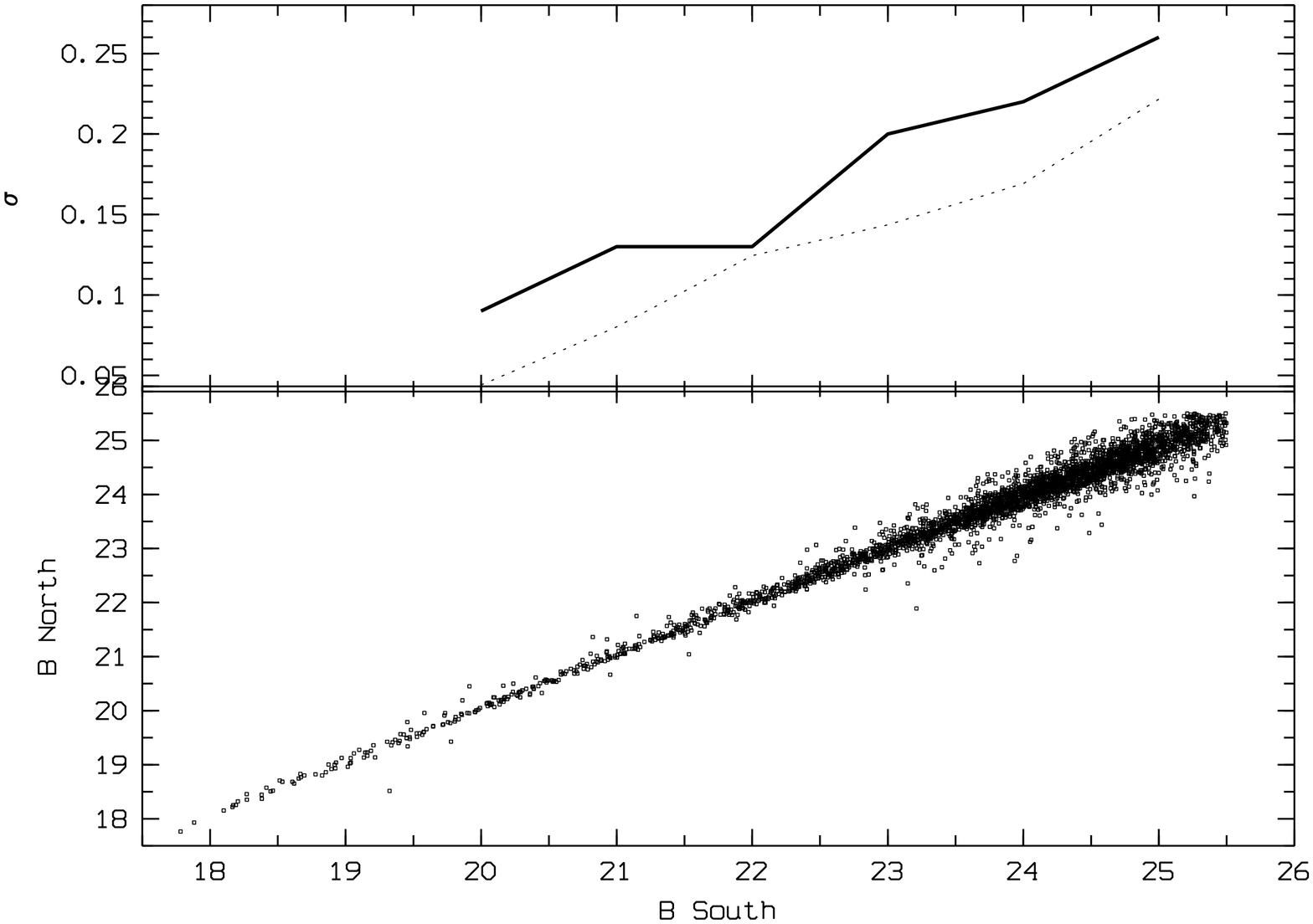,height=6cm,angle=-90}}
\caption[]{Upper graph: solid line: statistical 1$\sigma$
magnitude uncertainties between the North and South fields, dashed line:
quadratic sum of the internal error from SExtractor and of the
error due to the different seeings between different observations.
Lower graph: B North versus B South magnitudes.} \label{fit1}
\end{figure}

\begin{figure}
\centering \mbox{\psfig{figure=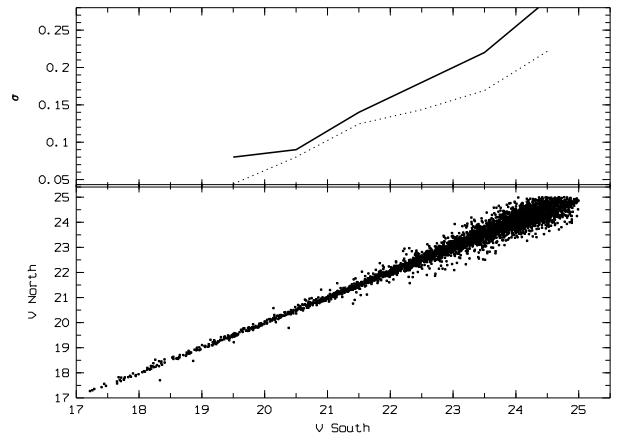,height=6cm,angle=-90}}
\caption[]{Upper graph: solid line: statistical 1$\sigma$ uncertainties
between North and South fields, dashed line: quadratic sum of
internal error from SExtractor and error due to the different
seeing between different observations. Lower graph: V North versus
V South magnitudes.} \label{fit2}
\end{figure}

\begin{figure}
\centering
\mbox{\psfig{figure=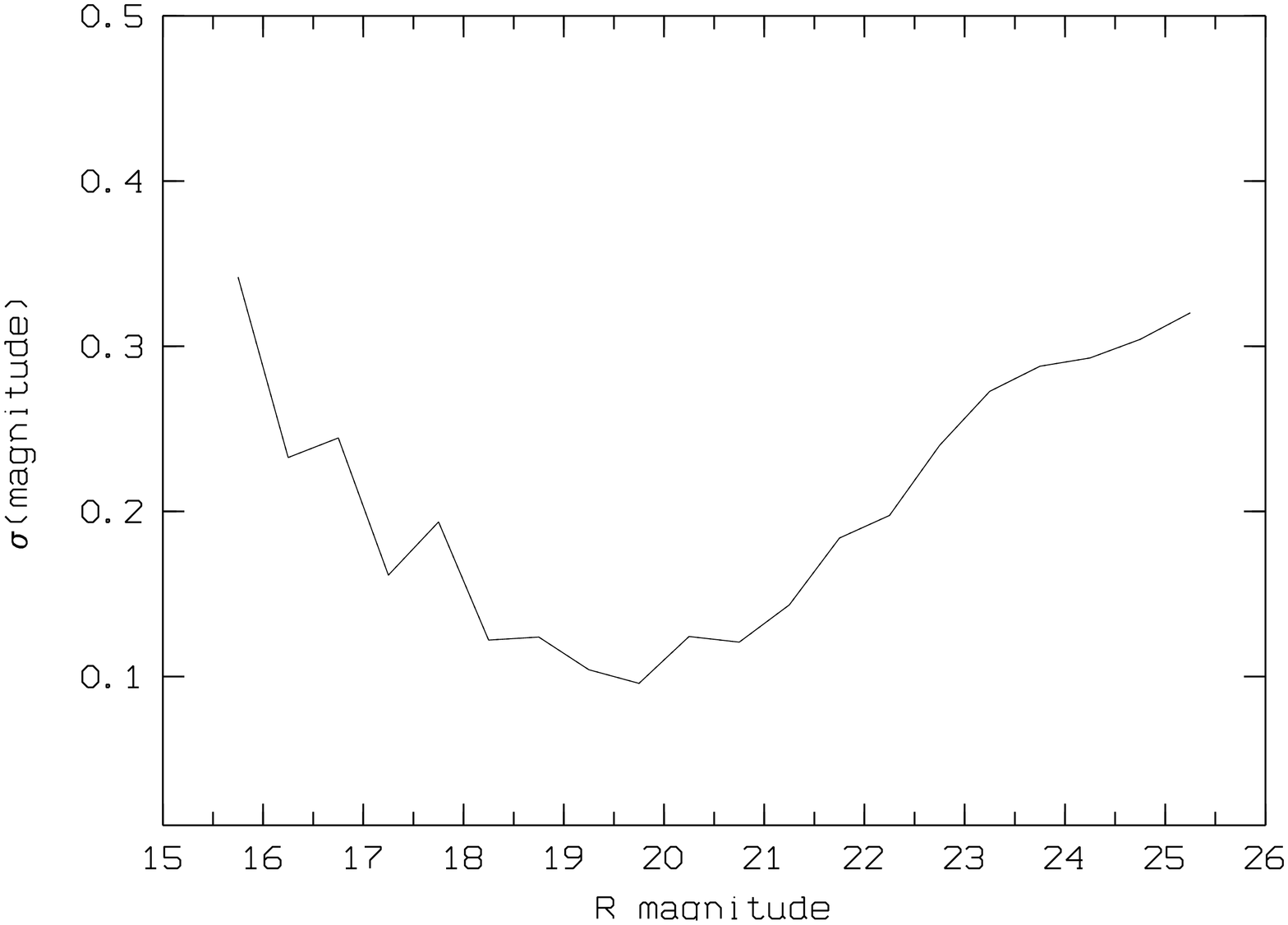,height=6cm,angle=-90}}
\mbox{\psfig{figure=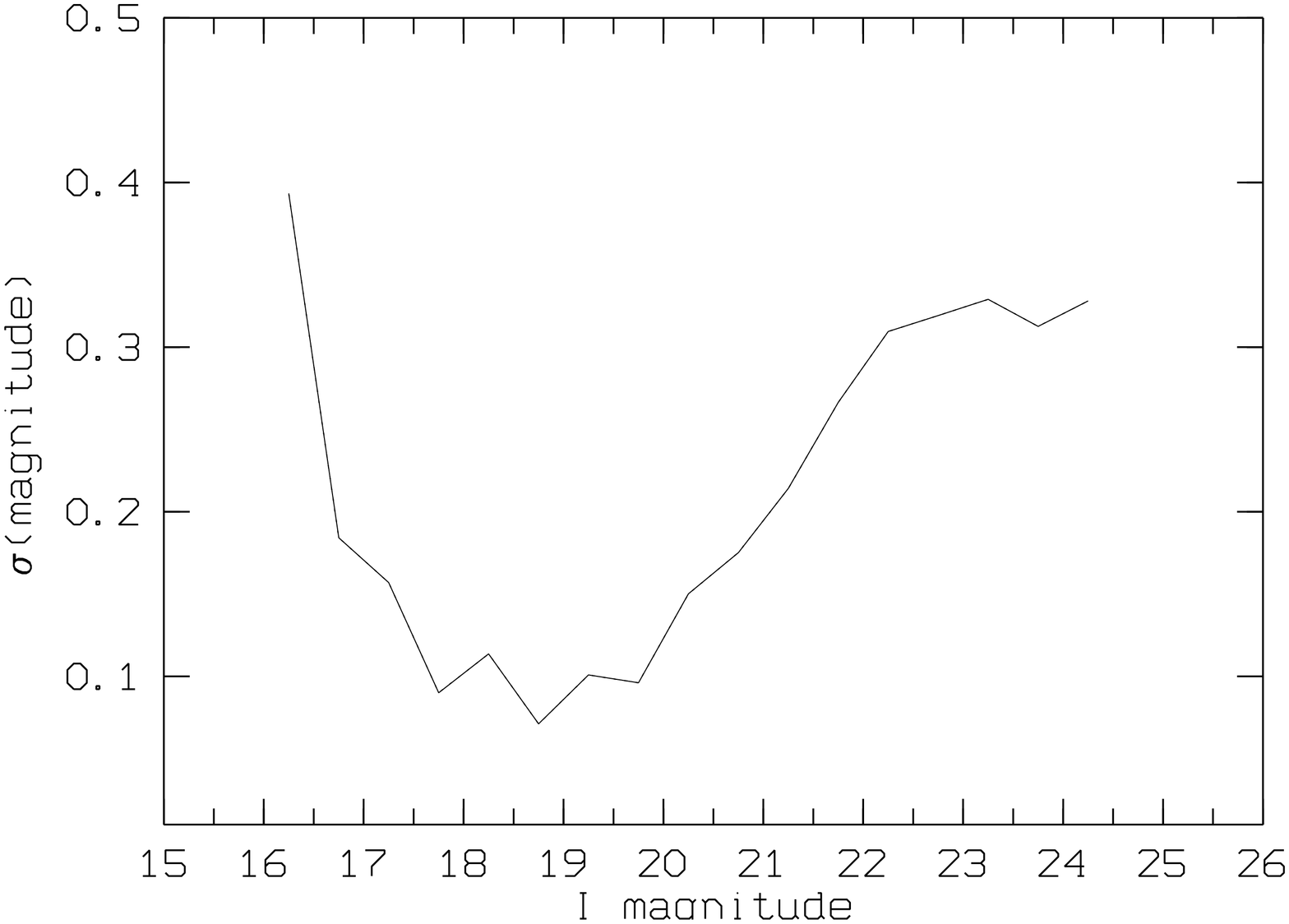,height=6cm,angle=-90}}
\caption[]{Upper figure: Statistical 1$\sigma$ uncertainty between first
epoch and second epoch data on South R field. Lower graph: Statistical
1$\sigma$ uncertainty between I band second epoch data in the common area.}
\label{fit3}
\end{figure}

To get an external estimate of the magnitude error budget, we used
the overlap between the North and South fields in the B, V and I
bands and the two South pointings in the R filter.
Results based on several thousand objects common to both fields
in the overlap
region for the B and V filters are shown in Figs.~\ref{fit1} and
~\ref{fit2} as a function of magnitude. Results of the two R observations
and in the common I area are shown in Fig.~\ref{fit3}.

The increasing uncertainty at the bright end in the R and
I bands is due to the second epoch data, because of a background
over-correction by Terapix near bright objects, and saturation effects
on deep images.

>From this analysis  we can approximate the magnitude
dispersion by regression laws for B, V, R and I filters with
parameters given in Table ~\ref{tab:fit}.

The photometric accuracy is better than 0.3 magnitude
in all bands, in good agreement
with the computed error budget. Also, the R error
estimate is probably an overestimate for the faint part
of second epoch data, which are clearly deeper
and of better quality.

\begin{table}
\caption{Fit of the magnitude uncertainty as a function of magnitude
deduced from the common areas. This
was used to compute uncertainties for individual galaxies: Uncertainty =
slope $\times$ magnitude + constant term.}
\begin{tabular}{llll}
\hline
Magnitude band & validity domain & slope & constant term \\
\hline
B       & [20,25.5] & 0.0340 & $-$0.5933 \\
V       & [20,24.5] & 0.0445 & $-$0.8352 \\
R       & [20,24] & 0.0093 & $-$0.0160 \\
I       & [20,23.5] & 0.0657 & $-$1.1772 \\
\hline
\end{tabular}
\label{tab:fit}
\end{table}

\subsection{Magnitude biases}

Systematic bias in the magnitudes of catalogued objects
could occur as a result of the presence of close neighbors
and saturation effects.
The effects of each of these is assessed below.

\subsubsection{Neighbor contamination}

To check this effect on the magnitude determination, we select the
objects flagged by SExtractor as having a chance of being biased by at
least 0.1 magnitude by a close neighbor.  The percentage of such
objects, measured in the R band, is shown in Fig.~\ref{stat} and
appears to be at a relatively constant level close to 15-20$\%$.  This
tells us that a relatively small fraction of our objects is
significantly affected by close neighbors.

\subsubsection{Saturation effects}

\begin{figure}
\centering \mbox{\psfig{figure=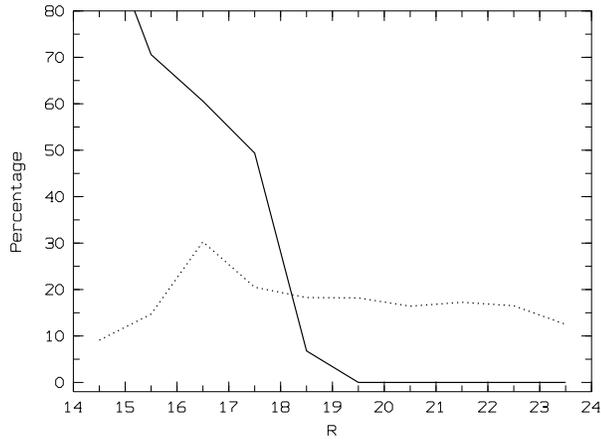,height=6cm,angle=-90}}
\caption[]{Solid line: percentage of objects with at least one
saturated pixel as a function of R magnitude. Dashed line: same
for biased magnitudes by 0.1.} \label{stat}
\end{figure}

In order to quantify the percentage of objects which could be
affected by saturation effects, we select objects flagged by
SExtractor as having at least one pixel saturated in their
profile. The result is given in Fig.~\ref{stat} as a function of
magnitude and shows that up to R=17.5 objects have a  probability
to have at least one saturated pixel higher than 50$\%$  . The
magnitude error is not quantified and depends on the object
brightness and morphology and so we chose to check the magnitude
of all brighter objects in shallower published data.

We have added published data only if they cover our whole field of view
(GP77, Terlevich 2001 and Komiyama 2002) to our
catalog of objects brighter than R=17.5 applying the mean magnitude shifts
computed in Section 4.7, without color corrections.

In the V band data, objects from the Terlevich et al. (2001) catalog
fainter than V$\sim$14 should not be saturated and can therefore cover the
whole range of our saturated objects.

In the B and R bands, data acquired by Komiyama et al. (2002) are supposed to not
be affected by saturation for magnitudes fainter than R=14. However, we
expect that very concentrated objects could still be saturated for
magnitudes somewhat fainter than R=14, and so we chose a conservative
approach by using the Komiyama et al. (2002) catalog only between
R=15.75 and 17.5 and the GP77 catalog for brighter objects.

In summary, all objects in the catalog fainter than R=17.5 come
from our CFH12k data. For objects between R=15.75 and 17.5, their
R and B magnitudes are taken
from Komiyama et al. (2002) or GP77 if the objects are classified as
saturated by us. Objects brighter than R=15.75 are taken from GP77 if
classified as saturated in our data. Finally, V magnitudes of objects
brighter than R=17.5 and classified as saturated are taken from
Terlevich et al. (2001).

In total we identified $\sim$540 potentially saturated objects in our catalog 
with R$<$17.5,
of which 272 are identified as galaxies in the literature.
The unidentified objects could be either stars (which are not
included in the Komiyama et al. and GP77 catalogs) or galaxies.
However, at least half of the R$<$17.5 objects are stars (see
Fig.~\ref{star_counts}), which are more easily saturated than galaxies,
and hence the largest part of the unidentified objects is likely to be stars.

We looked, however, at the 270 unidentified objects in our images and most
of them were obviously stars (slightly less than 70$\%$). A smaller subsample 
was made of
compact objects, possibly stars or spheroidal galaxies (slightly less than
30$\%$). Finally, a few galaxies were identified (7 galaxies).
We therefore decided to classify as stars all the
unidentified objects that were not obviously galaxies. This number of
$\sim$263 stars is, moreover, in good 
agreement with the Besan\c con model star counts (Gazelle et al. 1995).

\subsection{Magnitude comparison with other catalogs}

\begin{figure}
\centering
%\mbox{\psfig{figure=bsav_bgmptot.ps,height=6cm,angle=-90}}
%\mbox{\psfig{figure=rwest_rgmptot.ps,height=6cm,angle=-90}}
\mbox{\psfig{figure=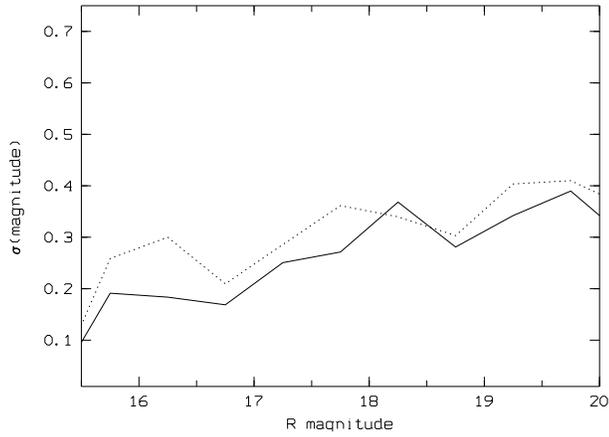,height=6cm,angle=-90}}
\caption[]{%Upper graph: B GP77 versus B CFHT. Middle graph: R GP77
%versus R CFHT. Lower graph:
Solid line: statistical 1$\sigma$ error between R GP77 and R
CFHT magnitudes, dashed line: B GP77 versus B CFHT magnitudes.}
\label{RGMP}
\end{figure}

 We first compared our catalog to the GP77 photographic
catalog (up to R$\sim$19.5) which covers a very large area that
encompasses our whole field. The result is given in
Fig.~\ref{RGMP}. The dispersion is lower than 0.4 magnitude at
R=20 and close to 0.1 in R and 0.15 magnitude in B at R=15.5, with
the following mean offsets:

B CFHT = B GP77 + 0.20

R CFHT = R GP77 + 0.52

\begin{figure}
\centering
%\mbox{\psfig{figure=vwest_vterltot.ps,height=6cm,angle=-90}}
\mbox{\psfig{figure=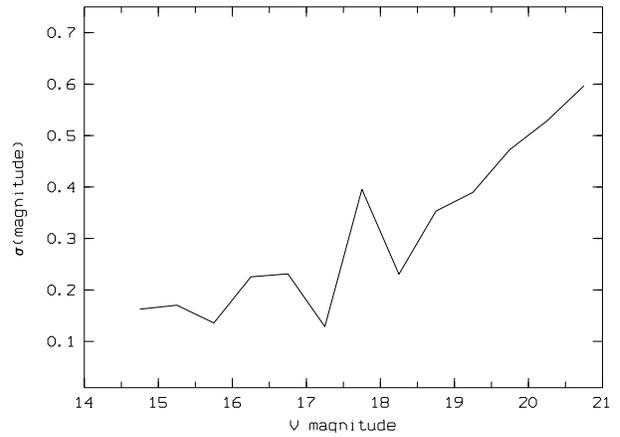,height=6cm,angle=-90}}
\caption[]{%Upper graph: V Terlevich et al. (2001) versus V CFHT.
%Lower graph: s
Statistical 1$\sigma$ uncertainties between V magnitudes from
Terlevich et al. and from our CFHT observations.}
\label{RT}
\end{figure}

\begin{figure}
\centering
%\mbox{\psfig{figure=bsav_bkomtot.ps,height=6cm,angle=-90}}
%\mbox{\psfig{figure=rwest_rkomtot.ps,height=6cm,angle=-90}}
\mbox{\psfig{figure=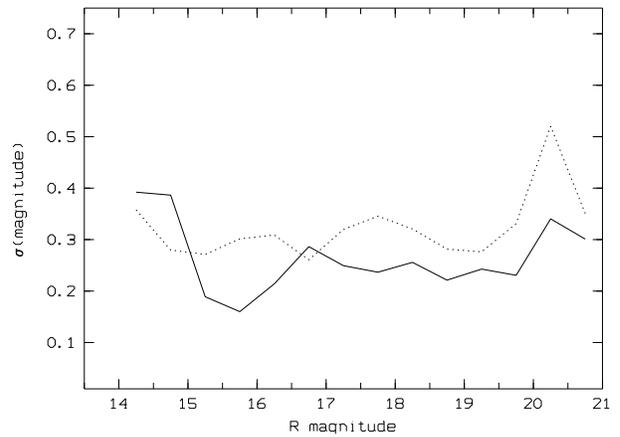,height=6cm,angle=-90}}
\caption[]{%Upper graph: B Komiyama versus B CFHT. Middle graph: R
%Komiyama et al. (2002) versus R CFHT. Lower graph: s
Solid line: statistical 1$\sigma$ uncertainties between R-band
magnitudes from Komiyama et
al. and our CFHT data, dashed line: statistical 1$\sigma$
uncertainties between Komiyama et al. and CFHT B magnitudes.}
\label{RK}
\end{figure}

We also compared our results with those of Terlevich et al. (2001) in
the V band and Komiyama et al. (2002) in the B and R bands. The
results are shown in Figs.~\ref{RT} and ~\ref{RK}.  B, V and R data
are in good agreement, with a dispersion at the bright end close to
0.2 magnitude for V, 0.25 magnitude for R and 0.3 magnitude for B. The
mean offsets are:

B CFHT = B Komiyama - 0.09

V CFHT = V Terlevich + 0.12

R CFHT = R Komiyama + 0.03

\begin{figure}
\centering
%\mbox{\psfig{figure=bsav_bb95tot.ps,height=6cm,angle=-90}}
%\mbox{\psfig{figure=rwest_rb95tot.ps,height=6cm,angle=-90}}
\mbox{\psfig{figure=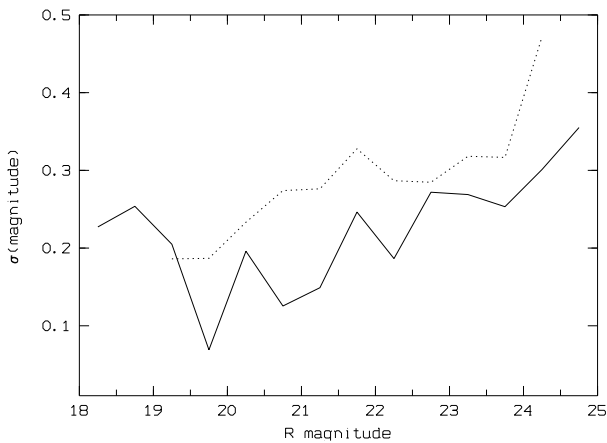,height=6cm,angle=-90}}
\caption[]{%Upper graph: B magnitudes from  Bernstein et al. (1995)
%compared to those from our CFHT data.
%Middle graph: R Bernstein versus R CFHT. Lower graph: s
Solid line: statistical 1$\sigma$ uncertainties between R Bernstein et
al. and R CFHT magnitudes, dashed line: statistical 1$\sigma$
uncertainties between B Bernstein et al. and B CFHT magnitudes.}
\label{RB95}
\end{figure}

Finally, we compared our results with the CCD observations of
Bernstein et al. (1995).  The result is shown in Fig.~\ref{RB95}. At
R=20, the dispersion is within 0.15 mag for the R band and 0.2 mag for
the B band. At the faint end of the catalog (R$\sim$25) the dispersion
is 0.35 in R and 0.45 in B.  At our catalog completeness level
(R$\sim$24) the dispersion is 0.25 in R and 0.3 in B. We note that
there is a moderate to strong difference in seeing between our data
and those of Bernstein et al., since the worst seeing for our data was
1.07~arcsec in B and 0.87~arcsec in R, while that of Bernstein et
al. was 1.31~arcsec. This probably explains why we see a rather smooth
variation of the dispersion between the Bernstein et al. and our B band
magnitudes, while the comparison in the R band shows visible oscillations.

The consistency between our catalog and others allows us to
merge some of the data to recover the saturated bright end of our
catalog. As we will use external data only for bright galaxies over
a rather limited magnitude range, and the mean offsets are small,
we do not include color terms from one photometric system to an other.

\section{Catalog completeness}

The catalog completeness has been measured in two ways:
by simulations and by comparison with a deeper catalog.

The simulation method adds artificial objects of different shapes
and magnitudes to the CCD images and then attempts to recover them
by running SExtractor again with the same parameters used for
object detection and classification on the original images. In
this way, the completeness is measured on the original images and
at different locations in the cluster. We investigated the catalog
completeness for point-like and faint low surface brightness
objects separately.  This is because part of the faint galaxy
population in clusters consists of faint low surface brightness
galaxies (R central surface brightness fainter than 24), which can
be nucleated (e.g. Ulmer et al. 1996). These objects are crucial
for understanding the cluster physics and, determining how deep we
can observe this population by comparison with point like objects,
is of major interest.

\subsection{Simulations for point-like objects}

The results at the 90\% mean completeness level are summarized in
column two of Table ~\ref{tab:comp} for all filters and in the two
fields.

\begin{figure}
\centering \mbox{\psfig{figure=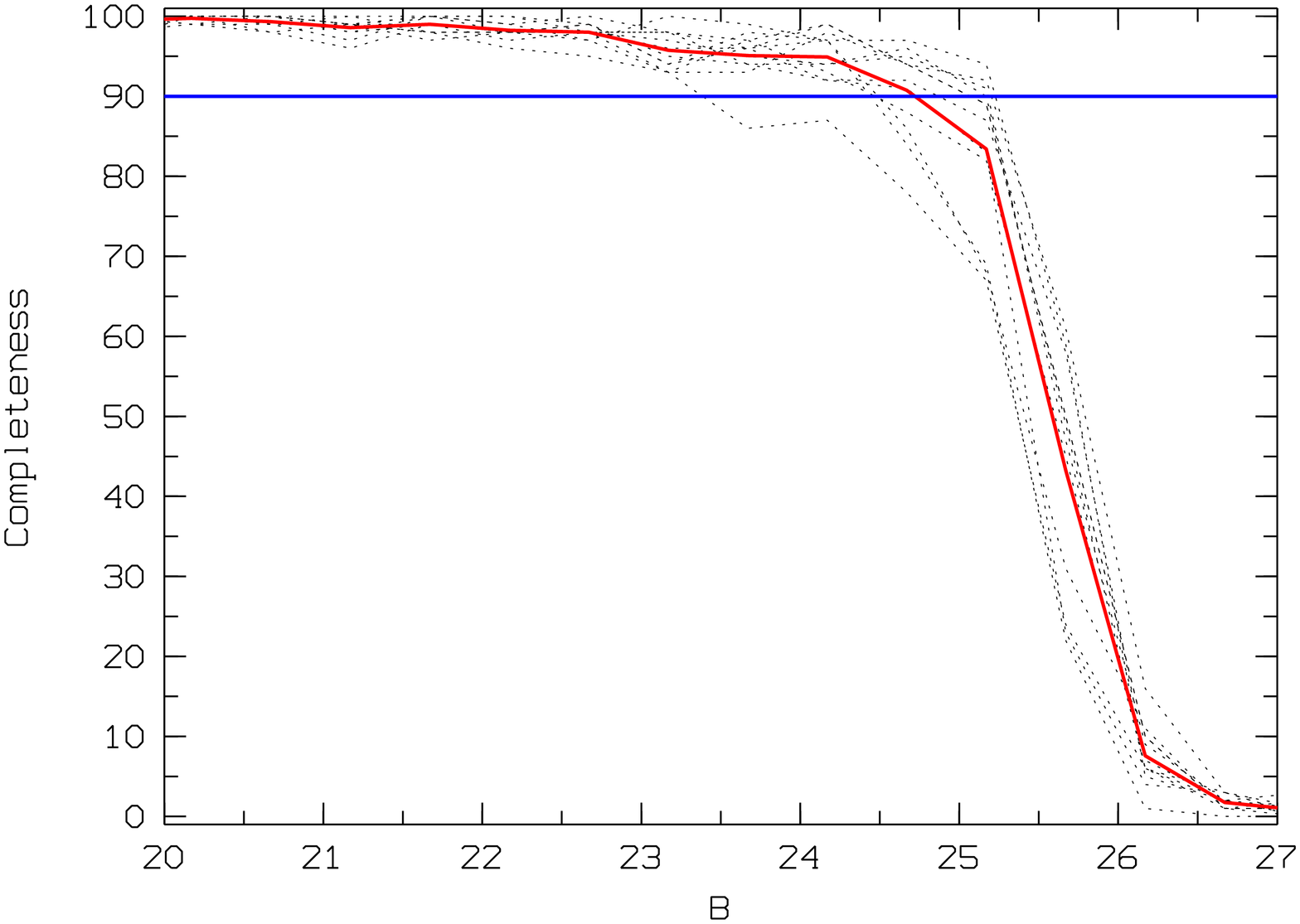,height=5.8cm,angle=-90}}
\mbox{\psfig{figure=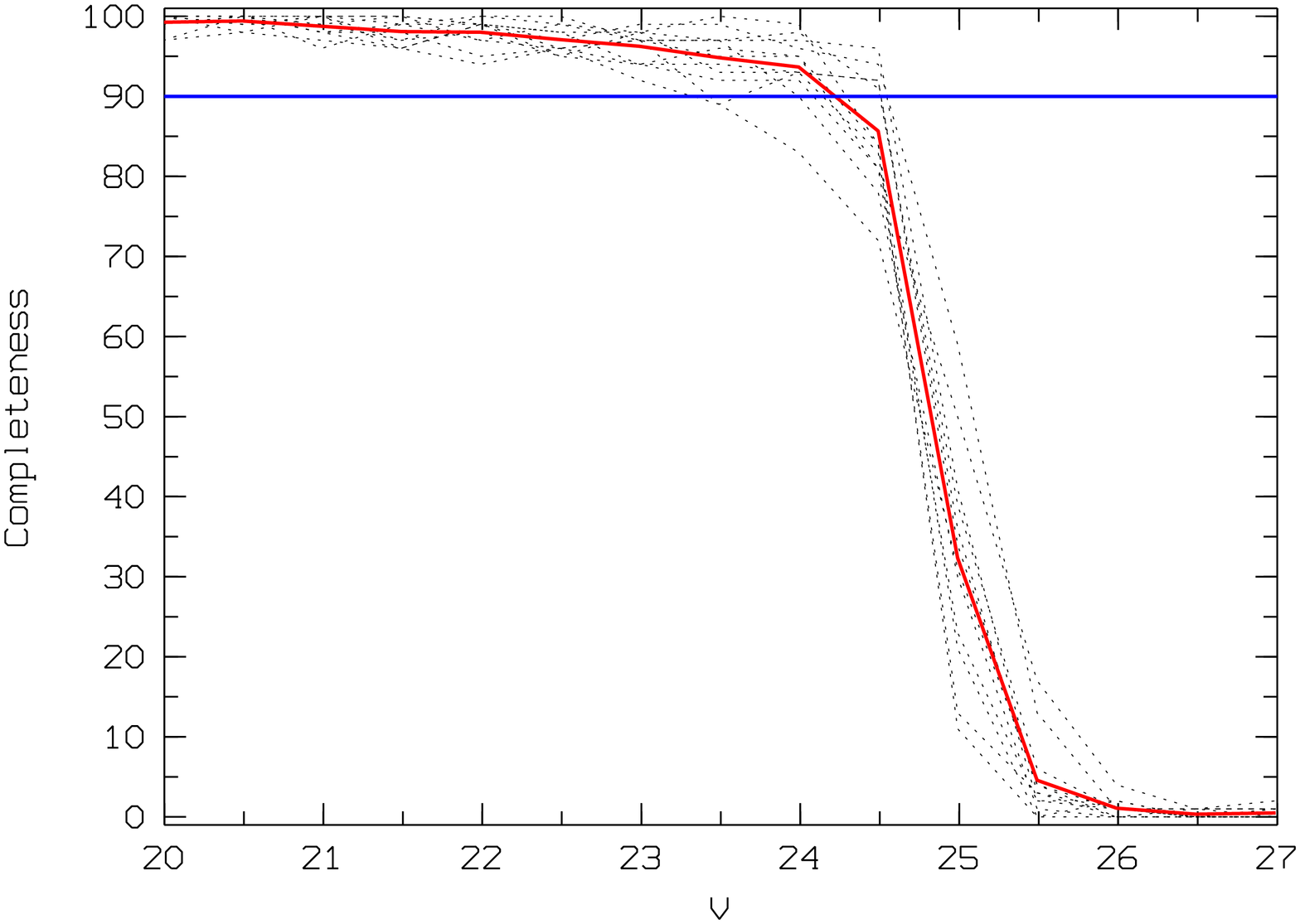,height=5.8cm,angle=-90}}
\mbox{\psfig{figure=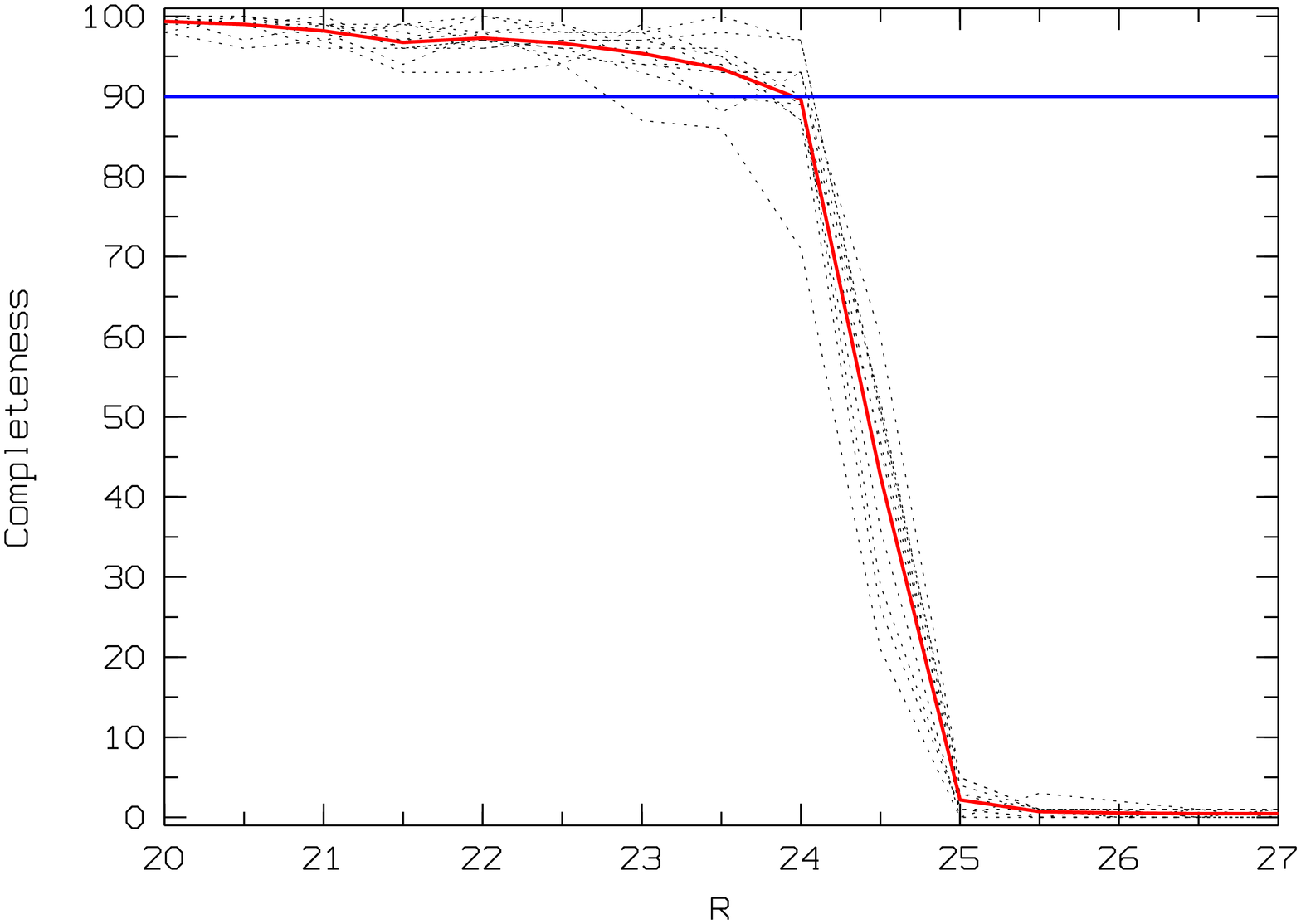,height=5.8cm,angle=-90}}
\mbox{\psfig{figure=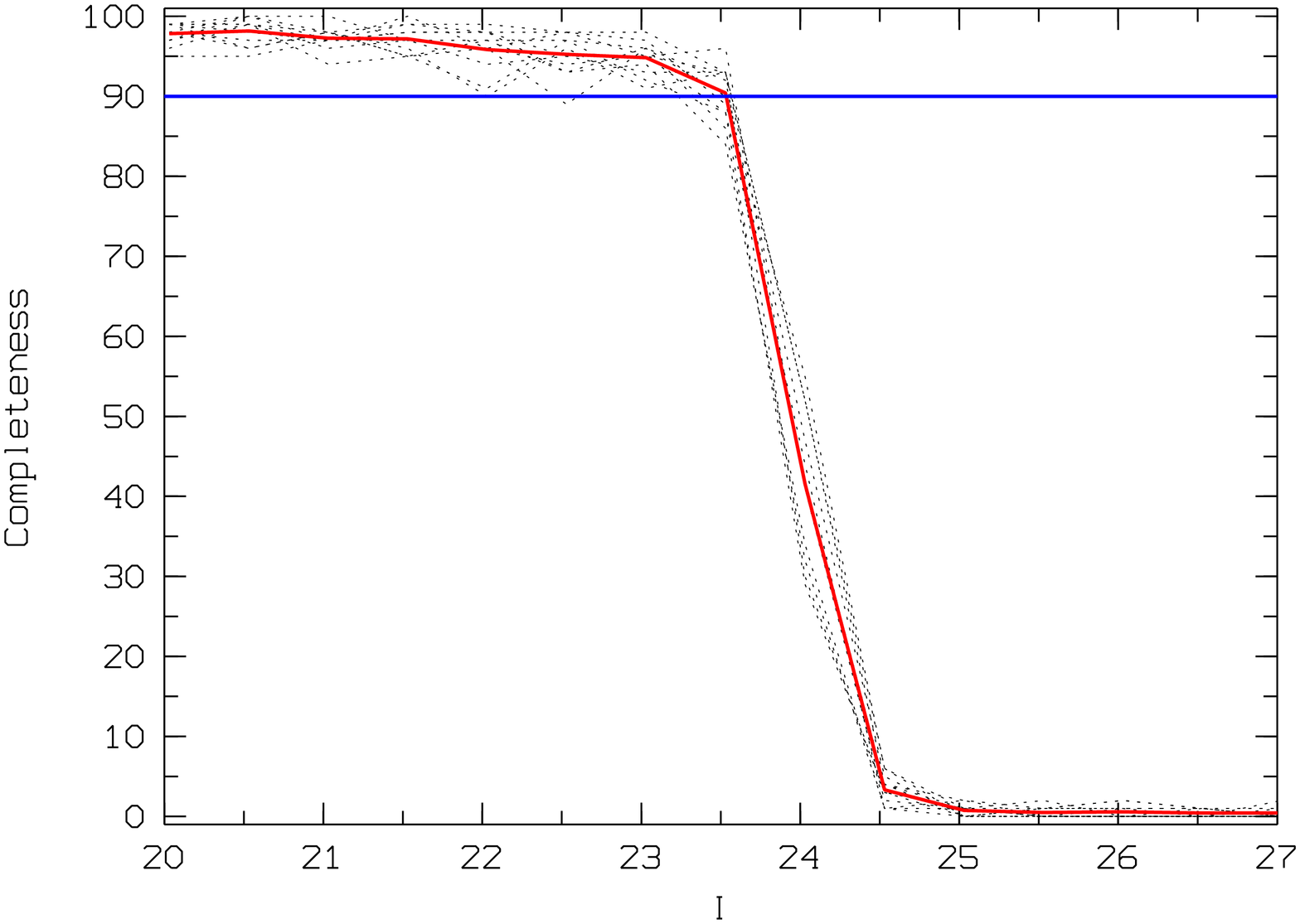,height=5.8cm,angle=-90}}
\caption[]{Completeness in percentages in B, V, R and I magnitudes
for point-like objects in the North field. The dotted lines show
the completeness of individual CCDs, the solid line is the mean
completeness for the North field and the solid horizontal line is
the 90\% completeness level.} \label{PL-N}
\end{figure}

An example is given in Fig.~\ref{PL-N}, which shows how the
completeness levels vary from CCD to CCD because of the QE variations
between individual CCDs and because of fluctuations in the
diffuse background light due to bright stars and
galaxies in the field.

\subsection{Simulations for low surface brightness objects}

\begin{figure}
\centering \mbox{\psfig{figure=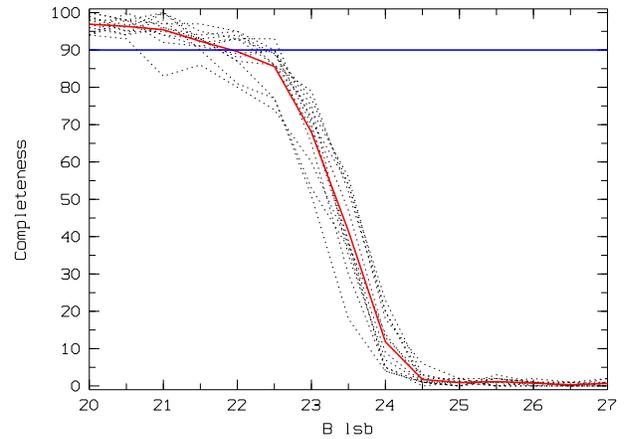,height=6cm,angle=-90}}
\caption[]{Same as Fig.~\ref{PL-N} for the faint low surface brightness
galaxies in the North field and in the B band.} \label{lsbPL-N}
\end{figure}

We estimated the completeness of our catalog for low surface
brightness galaxies using simulated point-like objects with a FWHM of
slightly more than 2 arcsec from a Gaussian profile. This is the
typical maximal size of a low surface brightness galaxy (e.g.  Ulmer
et al. 1996) in Coma. This method does not always take into account the true
profile of low surface brightness objects in Coma 
but is a good compromise between simulation simplicity
and result accuracy. Results are summarized in column 3 of
Table~\ref{tab:comp} and an example is given in Fig.~\ref{lsbPL-N} for
the B filter.

\begin{table*}
\caption{90\% and 50\% completeness levels for point-like and faint low surface
brightness galaxies.}
\begin{tabular}{lllll}
\hline
Magnitude & point-like 90\% & point-like 50\% & low surface brightness 90\% &
low surface brightness 50\% \\
\hline
B North      & B=24.75 & B=25.5 & B=22.0 & B=23.25  \\
B South      & B=25.   & B=26.   & B=21.75 & B=23.25 \\
V North      & V=24.25 & V=24.75 & V=21.0 & V=22.25  \\
V South      & V=24.   & V=24.5   & V=21.25 & V=22.25 \\
R North      & R=24.   & R=24.5   & R=20.75 & R=21.75 \\
R South      & R=24.5  & R=25.  & R=20.75 & R=22.5 \\
I North      & R=23.5  & R=24.  & R=20.5 & R=21.5  \\
I South      & R=23.25 & R=24. & R=20.25 & R=21.5  \\
\hline
\end{tabular}
\label{tab:comp}
\end{table*}

\subsection{Comparison with the Bernstein catalog}

Assuming that the Bernstein catalog is 100$\%$ complete at
our limiting magnitude, we can use it to obtain an
independent estimate of the completeness of our
catalog.  Based on this, we find 90$\%$ completeness at R=23.5, which
differs from that predicted by
our simulations, 90$\%$ at 24.5 for point-like objects and
50$\%$ at 22.5 for faint low-surface-brightness objects.

There are probably several reasons for the discrepancy.
First, faint galaxies are not exactly
stellar-like objects and their detection rate is, therefore, lower
than for stars. This probably explains part of the difference.

Second, the Bernstein (1995) field is
located very close to the two dominant galaxies of the Coma
cluster where there is a high level of diffuse light,
which can affect the detection of faint objects against
this background. The
Bernstein image was specifically treated to remove this light and
improve the detection rate of the faintest objects. As we
do not try to correct our data from diffuse light, this may contribute
to the difference in completeness limits.

\section{Star-galaxy separation}

Objects are classified as star-like or galaxies following two
criteria both based on the I band data. We used this band because
it was reduced in the most homogeneous way on the whole field and
was obtained under the best seeing conditions. The price to pay is
that not all objects detected in R have an I counterpart (R band
data are deeper than I band data).  However, as we will limit the
star-galaxy separation to I=21, this is not a serious concern as
most of the I$\leq$21 objects are detected in the 4 bands.

As a first criterion for star-galaxy separation, we use the SExtractor
class parameter,
which varies from 0 for galaxies to 1 for stars.  We classify as stars
objects with values greater than 0.98 (e.g. McCracken et al. 2003).  

\begin{figure}
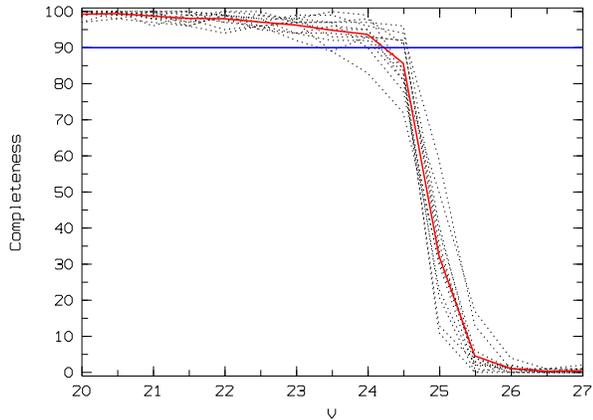

\centering
%\mbox{\psfig{figure=3810f131.ps,height=6.8cm,angle=-90}}
%\mbox{\psfig{figure=3810f132.ps,height=6.8cm,angle=-90}}
\caption[]{Central I surface brightness versus I total magnitude.
Crosses are objects identified as stars by SExtractor with
class parameter 0.98 or greater.
The heavy solid line shows the star-galaxy separation we
chose for the range I=[18.5;21]. Upper graph: North field. Lower
graph: South field.} \label{star_gal}
\end{figure}

The second criterion comes from the relation between the total
magnitude and the central surface brightness for different
types of objects. The stellar locus
clearly appears up to I$\sim$21 on Figs.~\ref{star_gal} and
~\ref{star_gal_zoom} showing unambiguously the good quality of the star-galaxy
separation up to this magnitude. The lines separation between stars and galaxies
were put in by hand in
order to optimally separate the stellar and galaxy loci.

For objects between I=17 and 18.5,  we did not use the total
magnitude / central surface brightness criterion because the
separation line starts to classify stars as galaxies around
I=18.25 and classify as stars all objects based on the SExtractor
class criterion. For objects between I=18.5 and 21 we used the
second criterion and classified as stars all objects below the
heavy solid line in Fig.~\ref{star_gal}.

Because the North and South data were obtained under different
seeing conditions, the criteria were
slightly different in the two fields and so we performed
the star-galaxy separation individually for each of the fields.

Beyond I=21, the star-galaxy separation is very unreliable, because of
possible confusion of stars with small galaxies (with seeing dominated
profiles) and we chose to classify all objects in this range as
galaxies.

\begin{figure}
\centering
%\mbox{\psfig{figure=3810f141.ps,height=6cm,angle=-90}}
%\mbox{\psfig{figure=3810f142.ps,height=6cm,angle=-90}}
\caption[]{Same as Fig.~\ref{star_gal} but in I=[18.8;22]. The
large squares in
 lower graph are the 3 faintest bona fide stars (determined by
external means) in the field.}
\label{star_gal_zoom}
\end{figure}

\subsection{Star counts: comparison to the Besan\c con model}

\begin{figure}
\centering
\mbox{\psfig{figure=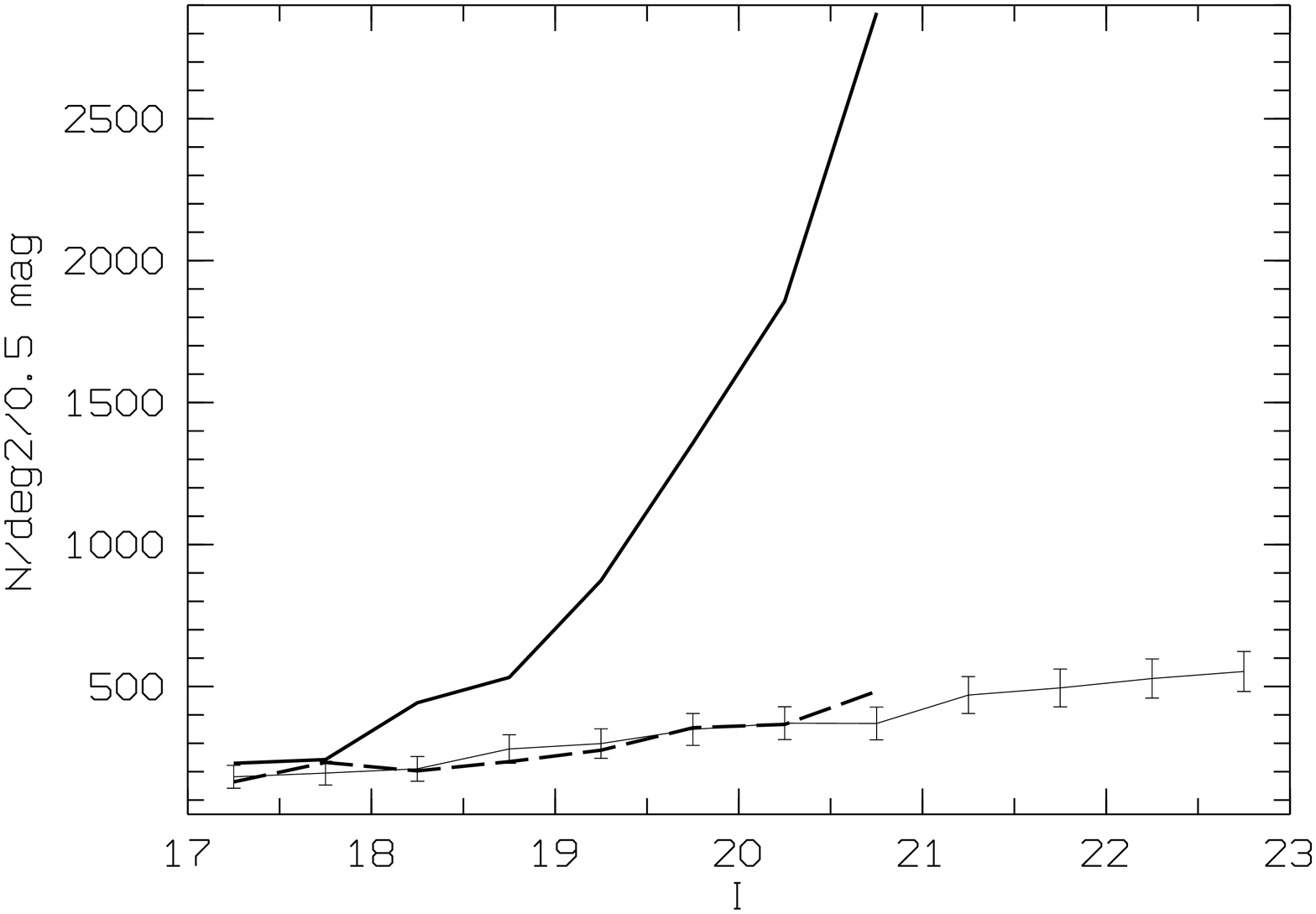,height=6cm,angle=-90}}
\caption[]{Star counts in our Coma catalog (long-dashed line),
star counts from the Besan\c con model (thin line with error bars)
and galaxy counts (solid line). Error bars are at the 3$\sigma$
level.} \label{star_counts}
\end{figure}

Our star counts shown in Fig.~\ref{star_counts} agree well
with the predictions from the Besan\c con model for our galaxy
(Gazelle et al. 1995 or http://bison.obs-besancon.fr/modele/) in the I
band and confirm the quality of the star-galaxy separation between
I=17 and I=21. At fainter magnitudes, the predictions show that the
contribution of stars is only 10\% at I=21 and drops quickly beyond
that.

\subsection{Moving objects and spectroscopically confirmed stars}

In order to further check the position of the stellar locus in the
surface brightness/total magnitude plots, it is useful to plot
known stars in the diagram. One way to do this is
to use proper motion to identify stars in
the Galactic halo.
However, a check of
the Vizier database at CDS revealed that
there are no published moving stars in the I=[19;21] range in this
field.

We therefore compared our images observed in 2000 to the DSS
photographic plate observed in 1955 and to the Bernstein et al. (1995)
field observed in 1991, in order to identify moving stars in our
fields.  Using a simple cross-correlation we select all CFHT objects
with a counterpart between 3.4~arcsec (i.e. $\sim$3 times the relative
astrometric precision between catalogs) and 15~arcsec in the DSS and
between 1.4~arcsec (i.e. 2 times the relative astrometric precision
between the catalogs) and 4~arcsec in the Bernstein et al
catalog. These upper values limit our search to proper motions smaller
than 400 mas/year, a reasonable upper limit (see e.g. the Tycho-2
catalog, Hog et al. 2000).  After examining each of these objects and
rejecting spurious detections, we were left with two bona fide moving
objects in the South field and none in the North field.  The first of
these is detected from the DSS plate
($\alpha_{2000}$=194.8629$^\circ$, $\delta_{2000}$=27.6834$^\circ$,
I=19.62). The second one is detected from the Bernstein et al. (1995)
image ($\alpha_{2000}$=194.8673$^\circ$,
$\delta_{2000}$=27.8909$^\circ$, I=19.67).  Additionally, observations
reported in Adami et al. (2000) led to the spectroscopic discovery of
a third star ($\alpha_{2000}$=194.8149$^\circ$,
$\delta_{2000}$=27.9257$^\circ$, I=19.05) in our fields.

These three objects are well located in the star locus, giving us
confidence in the reliability of our star-galaxy separation at least
down to I$\sim$20.

\section{Colors}

In order to check the reliability of the photometric spectral
energy distributions (SEDs) obtained in our fields, we have
compared the observed colors with theoretical expectations derived
for two different and well-defined samples of objects: stars and
galaxies with known redshifts. Although photometric redshifts and
star-galaxy identifications are beyond the scope of the present
paper, this section provides a reference on the quality of our SED
data from $\sim 4000$ \AA \ to 1 $\mu$m.

\subsection{Stars}

\begin{figure}
\centering
\mbox{\psfig{figure=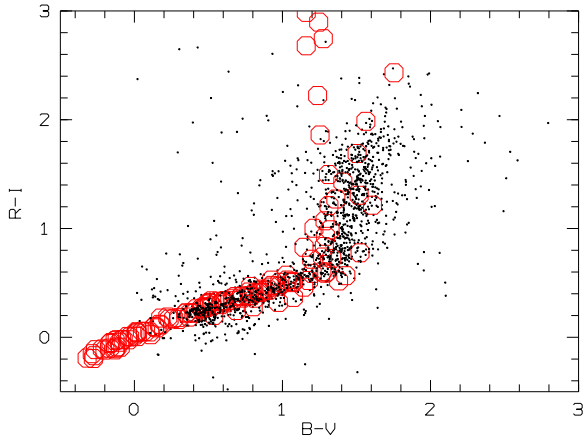,height=6cm,angle=-90}}
\caption[]{R-I versus B-V for I=[17;21] stars in our catalog
(dots), compared to synthethic values derived from the library of
Pickles (1998) (large open circles).} \label{star_locus}
\end{figure}

Synthetic colors for stars have been derived for a variety of
spectral types and luminosity classes, using the empirical stellar
library of Pickles (1998). We have adopted a detailed modeling for
filter transmissions, taking into account the total efficiency of
the system as a function of wavelength. Stars have been selected
in our catalogs according to the criteria given in Sect. 7
between $I$=17 and 21. There is a good agreement between
synthethic and observed colors of stars, as shown in
Fig.~\ref{star_locus}. On the color-color diagram $B$-$V$ vs.
$R$-$I$, both the theoretical locus of the observed main sequence
and the expected dispersion towards $B$-$V$ $\ge$ 1.2 are
well reproduced by observed stars.

\subsection{Galaxies}

The set of empirical templates compiled by Coleman, Wu and Weedman
(1980) was used to derive representative synthetic colors
for galaxies in our photometric system as a
function of redshift. The spectro-morphological types considered
are the following: E, Sbc, Scd and Im, thus providing a simple
representation of the galaxy population in the local universe.
Colors were obtained using the detailed filter transmissions
mentioned above for the CFH12K camera, with a simple
$k$-correction.

\begin{figure}
\centering
\mbox{\psfig{figure=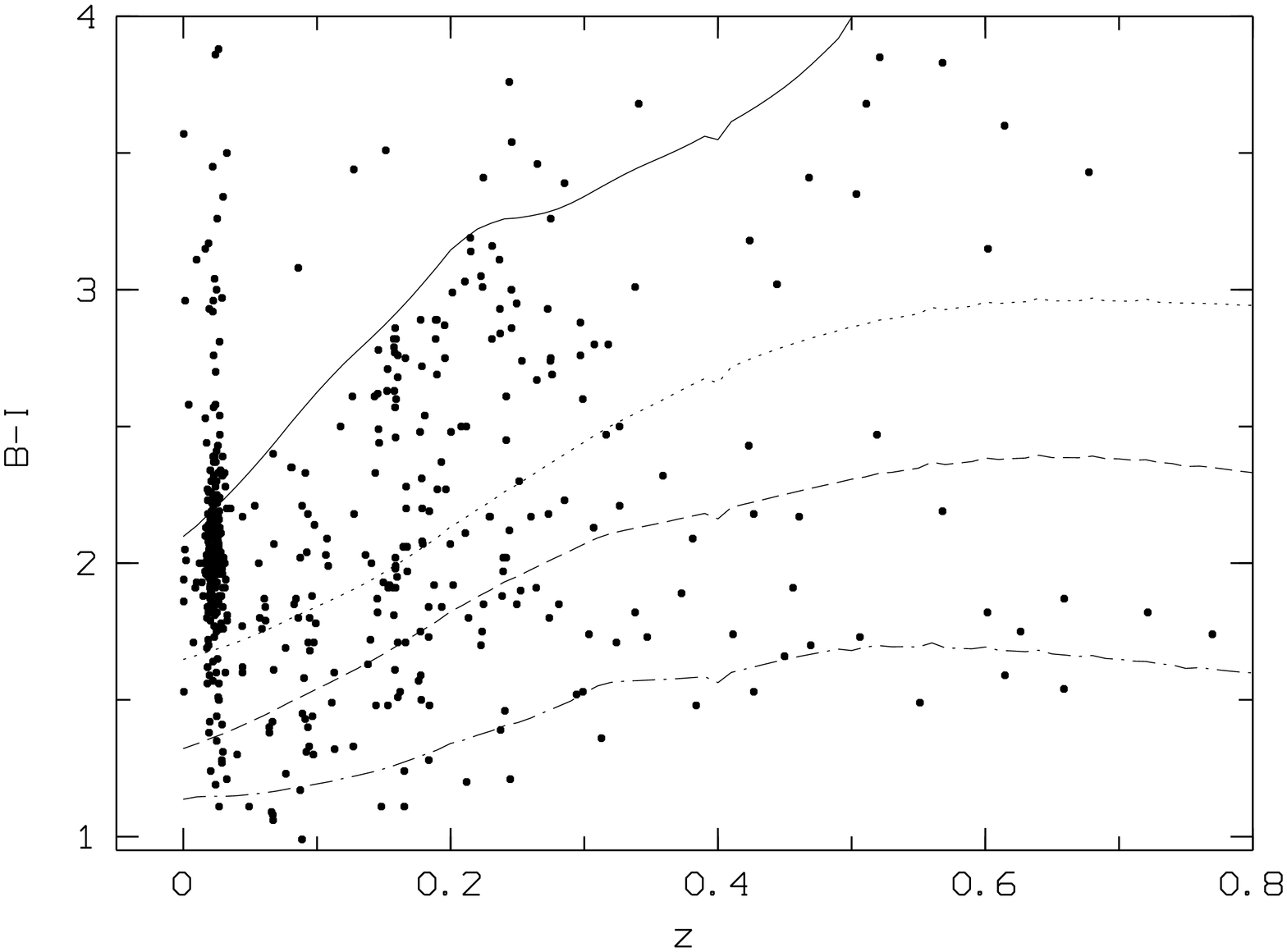,height=6cm,angle=-90}}
\mbox{\psfig{figure=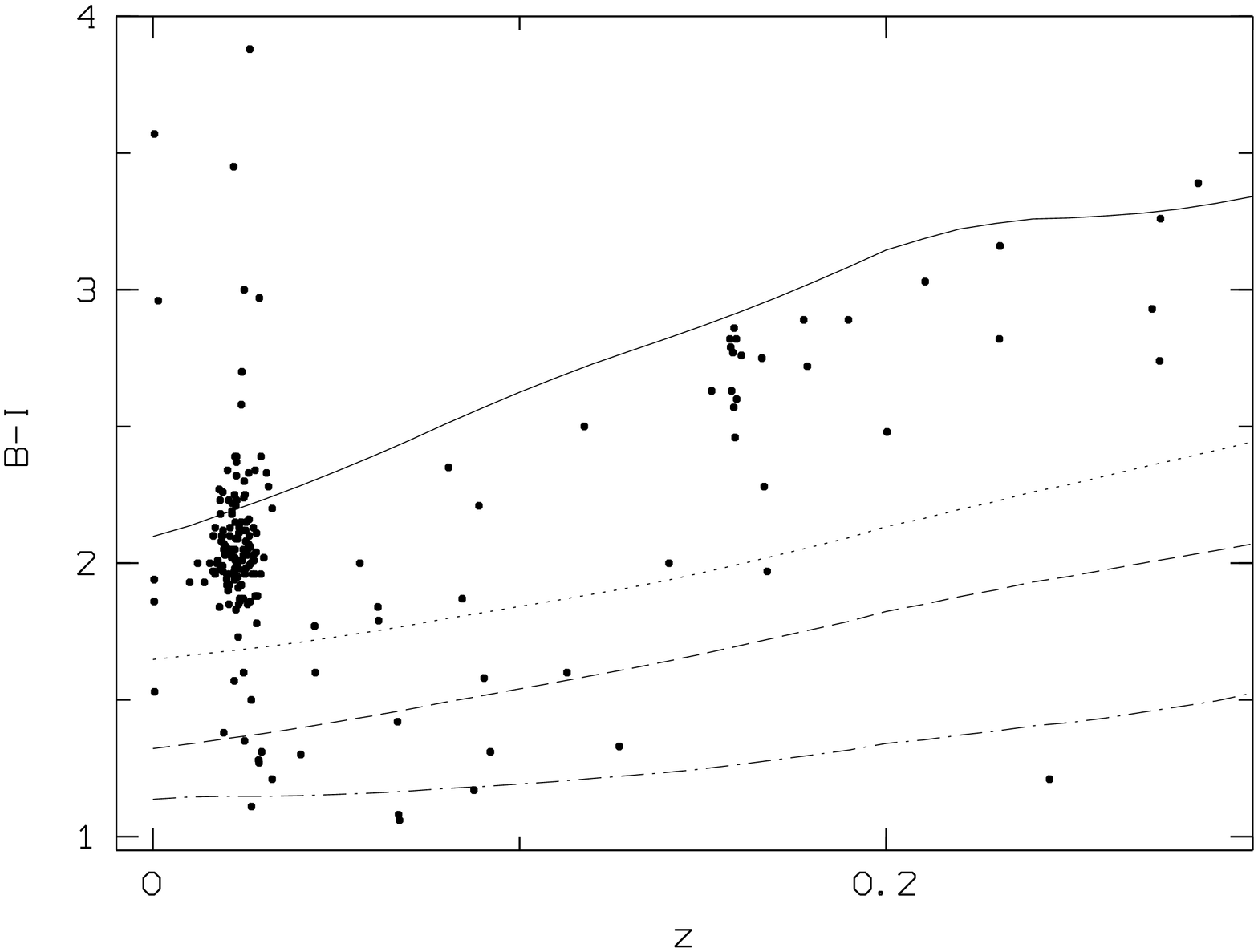,height=6cm,angle=-90}}
\caption[]{B-I versus redshift for a sample of galaxies with
spectroscopic redshifts in this field (from Adami et al. 2005b),
compared to simple model expectations derived from Coleman et
al. (1980) templates. From top to bottom: elliptical galaxy (solid
line), Sbc (dotted line), Scd (long dashed line) and Im (dot-dashed
line). Top panel: all galaxies. Bottom panel: galaxies visually
classified as early type galaxies according to their morphology. }
\label{galaxy_locus}
\end{figure}

A sample of 873 cluster members compiled from the literature (with
measured velocities between 4000 and 10000 km/s) is available, as
described by Adami et al. (2005b). Although this spectroscopic sample
is rather small, observed colors are found to be in good agreement
with model predictions. A representative example is given in
Fig.~\ref{galaxy_locus} for $B$-$I$ as a function of redshift. A large
majority of cluster galaxies display colors fully compatible with
early-type models, and the range of colors spanned by the whole
spectroscopic sample is in agreement with simple model
expectations. This is particularly true for galaxies morphologically
identified as early types from visual inspection (Biviano et
al. 1996).  It is worth noting that dust reddening has not been
considered here, and some objects displaying extremely red colors
would need this correction to fit into the scheme.
In particular, a moderate intrisic extinction A$_{V} \ge 0.5$ 
magnitudes (E(B-V) $ \ge$ 0.12, with a Calzetti et al. 2000 reddening law) 
provides B-V $ \ge$ 2.5 when applied to an Sbc template at the Coma 
redshift, using a simple dust-screen model. Thus, dust extinction 
could naturally explain the very red colors observed for a few galaxies 
(less than 10$\%$ of the total spectroscopic sample) in Fig.~\ref{galaxy_locus} 
compared to models. Also bright objects
with pixels close to the saturation limit could exhibit atypically
red colors.

Also bright objects
with pixels close to the saturation limit could exhibit atypically
red colors.

In summary, the bulk of the spectroscopic sample nicely
fits into the expected color-redshift diagrams, and this indicates a
good quality of the photometric SEDs for future studies based on
SED-fitting techniques.

\subsection{Red sequence in the Color Magnitude Relation}

\begin{figure}
\centering
%\mbox{\psfig{figure=3810f181.ps,height=6cm,angle=-90}}
%\mbox{\psfig{figure=3810f182.ps,height=6cm,angle=-90}}
\caption[]{Color magnitude relation in the Coma cluster. Small
dots: all galaxies on the Coma line of sight. Grey (B$\&$W
format) / green (color format) filled circles: Coma members. Black 
(B$\&$W format) / red (color format) filled circles: early type Coma
members. Black (B$\&$W format) / red (color format) solid line:
our fitted red sequence.  Upper figure: grey (B$\&$W format) /
green (color format) solid and dashed lines are the 1 and 3$\sigma$
error envelopes.  Lower figure: on an expanded scale, the solid grey
(B$\&$W format) / green (color format) line is the 1$\sigma$
error envelope and the two black (B$\&$W format) / red (color
format) dashed lines are the red sequences from Secker et
al. (1997) and Lopez-Cruz et al. (2004)}
\label{fig:CMRs}
\end{figure}

We also used the spectroscopic sample described in the previous
section to investigate the color magnitude relation (B-R versus R,
hereafter the CMR) in the Coma cluster and to compare our results with
published red sequences in Coma. The Coma cluster is known to exhibit a well
defined red sequence (e.g. GP77 or Mazure et al. 1988). The shape of this
relation is probably driven by metallicity effects in different mass
systems (e.g. Kodama $\&$ Arimoto 1997). Massive galaxies can be
redder than lower mass systems because they are able to retain more
metals and then to form more evolved stars. Early type galaxies in
this quite old cluster (e.g. Adami et al. 2005b) therefore form a well
defined sequence with a negative slope (faint and low mass early type
galaxies retain fewer metals than bright and massive ones and are
therefore bluer).

Fig.~\ref{fig:CMRs} shows this relation. We fit a red sequence using
only galaxies classified as early type galaxies and
likely Coma members based on their velocities
(between 3500 and 10000 km/s).
The relation is:

B-R = -(0.045$\pm$0.028) $\times$ R + (2.27$\pm$0.48)

The errors are quite large, but this is simply due to the small size of the
spectroscopic sample (even if one of the largest available for a cluster of
galaxies: Adami et al. 2005b) and to the uncertainties in the
visual morphological classification (potentially including late type
galaxies in the sample).  We also overplot on Fig.~\ref{fig:CMRs}
the error envelopes at the 1 and 3$\sigma$ levels around this mean
relation. Error envelopes were simply computed using the modeled
magnitude uncertainties (see Section 4.5) as a function of magnitude.

Several points are noteworthy.
First, the red sequence defined in this way is in good agreement with
the galaxy distribution in the R/B-R space. The high galaxy density
regions in this diagram are well correlated with the red sequence. We
also visually inspected the two early type galaxies inside the Coma
cluster with extreme B-R values. These two objects (around B-R=2.5,
R=16.3 and B-R=0.8, R=18.3) have several velocity measurements in the
literature
(5 and 4 respectively), all consistent with them being Coma members
(7451 km/s and 8043 km/s). These objects must certainly have had peculiar
histories that made their B-R colors atypical. They probably experienced
dynamical encounters at least for the reddest one which has a peculiar
shape in our images. However, discussing the precise evolution of these
  two objects is beyond the scope of this paper.

Second, the galaxies inside the Coma cluster but without morphological
information available in the literature (filled grey circles in
Fig.~\ref{fig:CMRs}) are also very close to this red sequence,
defining a relatively narrow relation down to R$\sim$19 (consistently with
e.g.  Adami et al. 2000). What occurs at fainter magnitudes remains
uncertain as we do not have deep enough spectroscopy to reach any
conclusion (see also
Adami et al. 2000).

Finally, our red sequence agrees very well with results published
previously in the literature, as shown in Table ~\ref{tab:cmrlit}.
Some marginal differences exist for the constant terms but are due
to slight differences in the B and R filters used by different
studies.

Investigations of the faint part of the red sequence are beyond the
scope of this paper and will be addressed in a future work.

\begin{table}
\caption{Literature red sequences in the B-R/R sequence.}
\begin{tabular}{lll}
\hline
Authors & Slope & Constant term \\
\hline
present paper & -0.045 &  2.27 \\
Gladders et al. (1998) & -0.045 & - \\
Lopez-Cruz et al. (2004)& -0.046 &  2.22 \\
Secker et al. (1997) & -0.056 &  2.41 \\
\hline
\end{tabular}
\label{tab:cmrlit}
\end{table}

\section{Summary and prospects}

We have presented and discussed the properties of a catalog of more
than 60,000 objects in the Coma cluster based on deep CFHT observations
in the B, V, R and
I bands and spanning an unprecedented 10 magnitude range in one of the
largest $\sim$0.72$\times$ $\sim$0.82 deg$^2$ areas presently
available for Coma at this depth. The catalog is complete
at the 90$\%$ level down to R$\sim$24 for stellar-like objects and
to R$\sim$20.75 for faint low surface brightness (see Sect.~5.2) galaxy-like
objects.

Astrometry is accurate to 0.5 arcsec, except for first epoch data on
the edges and corners of the CCDs. Magnitude errors are smaller than
$\sim$0.3 in all bands down to R=25. Saturation effects on objects
brighter than I=17.5 have been removed by using shallower published
data. More generally, galaxy magnitudes are in good agreement with
published data and the colors of galaxies and stars are in good agreement
with synthetic models. The bright part of the CMR agrees well with that
derived for galaxies with known redshifts in the Coma cluster, and with
previous CMRs published in the literature.

The star-galaxy separation is robust for all objects brighter than
I=21, and the star counts fit the Besan\c con model very well.

These data have already been used to search for diffuse emission in
the Coma cluster (Adami et al. 2005a) and to look for and analyze
properties of faint low surface brightness galaxies (Adami et
al. 2006).  We plan next to use these data to investigate the
properties of the different cluster galaxy classes and to derive the
luminosity functions of the Coma cluster galaxies in various bands and
in various regions of the cluster. This should deepen our knowledge of
environmental effects on galaxy luminosity functions, already shown to
be strong in Coma (e.g. Lobo et al. 1997).

The catalog described above will be made public
by the CENCOS center at http://cencosw.oamp.fr/. Individual catalogs for
the second epoch R and I band data will also be available upon request.

\begin{acknowledgements}
The authors thank the referee for useful and constructive comments.
We also thank the CFHT and Terapix teams, especially
Mireille Dantel-Fort for reducing the second epoch data and the
French PNG for financial support.
M.J. West acknowledges support from U.S. National Science Foundation grant
AST-0205960.
\\
\end{acknowledgements}

\end{document}